\documentclass[reprint,amsmath,amssymb,aip,jcp,author-numerical]{revtex4-1}
\usepackage{graphics}
\usepackage{graphicx}
\usepackage{dcolumn}% Align table columns on decimal point
\usepackage{bm}
\usepackage{natbib}
\usepackage{hyperref}
\usepackage{color}
\usepackage{tikz}

\begin{document}
\title{Mitochondrial cristae modeled as an out-of-equilibrium membrane driven by a proton field}
\author{Nirbhay Patil}
\affiliation{Sorbonne Universit{\'e}, CNRS, Laboratoire de Physique Th{\'e}orique de la Mati{\`e}re Condens{\'e}e (LPTMC, UMR 7600), F-75005 Paris, France}
\affiliation{Sorbonne Universit{\'e}, CNRS, Institut de Biologie Paris-Seine, Laboratoire Jean Perrin (UMR 8237), F-75005 Paris, France}
\author{St\'ephanie Bonneau}
\affiliation{Sorbonne Universit{\'e}, CNRS, Institut de Biologie Paris-Seine, Laboratoire Jean Perrin (UMR 8237), F-75005 Paris, France}
\author{Fr\'ederic Joubert}
\affiliation{Sorbonne Universit{\'e}, CNRS, Institut de Biologie Paris-Seine, Laboratoire Jean Perrin (UMR 8237), F-75005 Paris, France}
\author{Anne-Florence Bitbol}
\affiliation{Sorbonne Universit{\'e}, CNRS, Institut de Biologie Paris-Seine, Laboratoire Jean Perrin (UMR 8237), F-75005 Paris, France}
\affiliation{Institute of Bioengineering, School of Life Sciences, {\'E}cole Polytechnique F{\'e}d{\'e}rale de Lausanne (EPFL), CH-1015 Lausanne, Switzerland}
\author{H\'el\`ene Berthoumieux}
\affiliation{Sorbonne Universit{\'e}, CNRS, Laboratoire de Physique Th{\'e}orique de la Mati{\`e}re Condens{\'e}e (LPTMC, UMR 7600), F-75005 Paris, France}

\begin{abstract}
As the places where most of the fuel of the cell, namely ATP, is synthesized, mitochondria are crucial organelles in eukaryotic cells. The shape of the invaginations of the mitochondria inner membrane, known as cristae, has been identified as a signature of the energetic state of the organelle. However, the interplay between the rate of ATP synthesis and the crista shape remains unclear. In this work, we investigate the crista membrane deformations using a pH-dependent Helfrich model, maintained out-of-equilibrium by a diffusive flux of protons. This model gives rise to shape changes of a cylindrical invagination, in particular to the formation of necks between wider zones under variable, and especially oscillating, proton flux.
\end{abstract}

\maketitle

 %are the place of both the cemical reaction take place and where the protomotrice force is generated. 

 %che complex interplay between diffusion, chemistry, and membrane dynamics is not understtof yet}
\section{Introduction}
Mitochondria are important organelles of eukaryotic cells often called the “powerhouses of the cell”, due to their role
in the synthesis of Adenosine Tri-Phosphate (ATP) from
Adenosine Di-Phosphate (ADP) and an inorganic Phosphate (P$_i$). These organelles of micrometric size comprise an inner membrane (IM), which delimits a region called the matrix, and an outer membrane (OM) \cite{frey2000,mannella2006structure}. The volume between the IM and the OM is called the intermembrane space (IMS). The inner membrane presents numerous tubular invaginations of nanometric size, called cristae, where ATP synthesis takes place. The liquid inside the cristae is isolated from the IMS by the cristae junction, an aggregate of proteins that limit the diffusion\cite{koob2014novel}. Recently, it has been shown that cristae have a higher membrane potential than the intervening boundary membranes, involving confined proton loops and individual functioning of each cristae within the same mitochondria \cite{wolf2019}. It has been observed experimentally in isolated mitochondria that the cristae assume different shapes depending on the state of ATP production. Five stationary states (State I to State V) have been introduced to describe the energy status of isolated mitochondria \cite{chance55}, but most present research focuses on State III and State IV, since it allows to mimic the in vivo situation where an increase of energy demand and energy production occurs. Here, we will only consider these two states.   A high rate of proton injection by the respiratory chain and of ATP production  (State III) is associated with bumpy and wide cristae tubules, while a low rate of proton injection and ATP production State IV) is associated to a more regular cylindrical shape  as illustrated by Fig. 1. Moreover, recent experiments employing super-resolution imaging techniques have directly evidenced the dynamical deformations of cristae in cultivated cells\cite{wang2019}.
\begin{figure}
\includegraphics[scale=0.4]{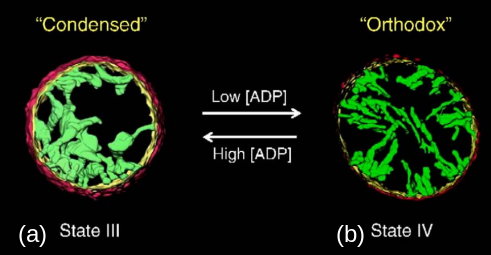}
\caption{Cross section of a mitochondrion in different states
of ATP production\cite{manella2006}. (a) State III, for which sugar in excess is available leading to a high rate of ATP production. (b) State IV, for which no sugar is available leading to a vanishing ATP production rate. 
Mitochondria (left to right) have diameters of 1500 nm and 500 nm.}
\end{figure}

The endothermic reaction ADP +P$_i$ $\rightarrow$ ATP , is catalyzed by the ATP-synthase which is located in the curved zone of the crista membrane\cite{cogliati2016mitochondrial}. Traditionally, the ATP synthase enzyme was supposed to use the bulk proton electrochemical potential gradient, involving both  the bulk pH and the electric potential difference between the cristae and the matrix, as energy supply \cite{mitchell1961coupling}. More recently, it has been established that the proton flux going down the gradient and allowing the rotor to turn with respect to the stator in the ATP synthase enzyme  is probably localized on the surface of the crista membrane \cite{gennis2016, morelli2019}. This flux is established and maintained by the respiratory chain which injects protons from the matrix on the IMS part of the crista membrane. The proteins of the respiratory chain are located in the weakly curved zones of the invagination and thus are spatially separated from the ATP-synthases. Recent {\it in vivo} pH measurement show that the pH decreases along the crista membrane between the proton source (the respiratory chain proteins) and the proton sink (the ATP synthase)\cite{rieger2014lateral}.  

Membrane deformation driven by out-of-equilibrium chemical dynamics is an ubiquitous phenomenon in living cells. A mechanism of hydro-osmotic instabilities generated by ion pumps  has been recently suggested to describe the dynamics of the contractile vacuole complex \cite{al-izzi2018}. Surface deformation driven by diffusion of an `active' species is commonly observed {\it in vivo} such as the division of eukaryotic cells by accumulations of myosin motors at the cell ring\cite{turlier2014}. A phenomenological model has been proposed to study the coupling between diffusion of active agents and surface shape by introducing a modified Helfrich model associated with an active tension coupled to the 2-dimension diffusion of the chemical regulator on the deforming surface \cite{mietke}. Note that the coupling between a diffusive active agent and the bending of a membrane has not been included in the model \cite{berthoumieux2014}.

The dynamical coupling between crista shape and ATP production rate is a recent discovery and the physicochemical mechanism  at the origin of this coupling still needs to be characterized. The lipid composition of the IM has been pointed out as a key point for the ATP-synthesis machinery. Indeed, the crista membrane is enriched in cardiolipin, and loss of mature cardiolipins affects the shape of the cristae and perturbs its function \cite{acehan2011}. These lipids possess a protic hydrophilic head and {\it in vitro} experiments have shown that tubular invaginations can be created by an externally controlled pH gradient in giant vesicles comprising cardiolipins \cite{khalifat2008membrane}. A theoretical description modeling these giant vesicles as locally planar bilayer membranes with lipid density and composition heterogeneities in each monolayer \cite{AF} has successfully reproduced the dynamics of the membrane in the regime of small deformations \cite{AF2}. In this model, composition can represent e.g. the acid and the basic form of cardiolipins, which is controlled by the local pH field.  However, because the crista membrane is enriched in proteins, representing up to 50 per cent of its mass, a detailed model describing a pure lipid bilayer and including the slippage between the two monolayers may not be necessary to describe this system. Therefore, here  we consider a simpler and more phenomenological Helfrich model with pH-dependent parameters.

This work proposes a model for the dynamics of the deformation for the crista membrane between state IV and state III.  We start with a reaction-diffusion system describing the proton flux on a cylinder (representing the crista membrane) which contains a proton source, a proton sink and a reflecting barrier. The resulting proton concentration field will be considered as the driving force inducing the membrane deformation. We then propose a pH-dependent Helfrich model,  in which the bending modulus, spontaneous curvature and tension depend on the local proton concentration, assuming small variations of this concentration. We derive the Green function of the system and study the phase diagram of the crista shape in this model. Finally we solve the hydrodynamic equations of the system for a proton field oscillating between the state III and the state IV and show that such a model generates dynamical deformations of the membrane, as well as the formation of necks and bumps, along the cylinder leading to a rougher surface in state IV. The last part is devoted to the conclusion.

\section{Model of a mitochondrial crista}
\subsection{Proton field along the crista}
We model the crista as an axisymmetric cylinder of membrane of finite length $L$ closed by a spherical cap (see Fig. 2). In experimental observations\cite{davies2011}, cristae feature different shapes, the most common being an elongated pancake, with rows of ATP synthase situated at the rim of the protrusion and respiratory chain proteins located in the flat zone. We nevertheless chose to work in the cylinder geometry, which is a simple special case of this pancake shape. Indeed, this geometry is analytically tractable, and despite its simplicity, it captures several key ingredients of the system, such as the nanometric confinement and the presence of zones of various curvatures. The protons diffuse on the crista surface at the concentration 
\begin{equation}
[{\rm H}^+](s)=[{\rm H}^+]_{\rm IV}+h(s),\label{eq:defn_h(s)}
\end{equation}
where  $s$ is a coordinate parameterizing the position along the tube, while $[{\rm H}^+]_{\rm IV}$ represents the constant concentration in state {\rm IV} taken as a reference, and $h(s)$ is the variation in the proton concentration induced by the functioning of the respiratory chain and the ATP synthases.  Note that the concentration is expressed as a number of protons per unit of length, taking advantage of the one-dimension symmetry of the problem. At one end of the cylinder, $s=0$, one finds a reflecting barrier for the protons modeling the cristae junction, while at the other end, $s=L$, a ring-shaped proton sink models the ATP synthases. Between the two, at $s=L_s$, a ring-shaped proton source models the respiratory chain. 
This model for the geometrical confinement of the ATP synthesis machinery takes into account the spatial separation of the proton source and sink and their respective localization in zones of low and high curvature \cite{davies2011}.

 \begin{figure}
\includegraphics[scale=0.25]{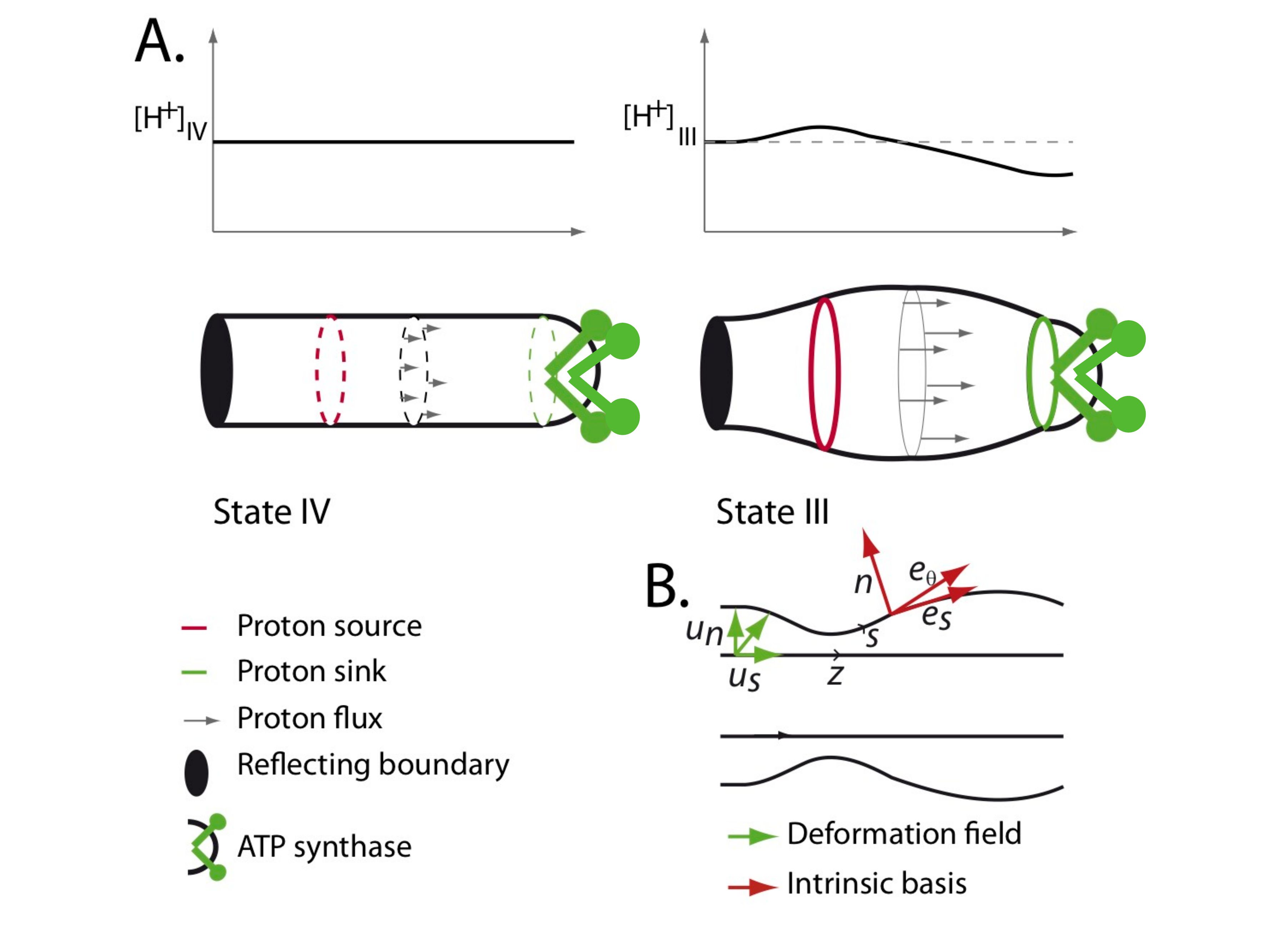}
\caption{Schematic representation of a crista. {\bf A.} The plots represent the proton concentration on the surface in state III and IV. The tubes represent the shape of the invagination in state III and IV. {\bf B.} Deformation fields of the membrane and intrinsic basis of the deformed surface.}
\end{figure}

We assume that $h(s)$ is small, {\it i.e.}  $h(s)/[{\rm H}^+]_{\rm IV} \ll 1$, and that the tube shape does not deviate much from a regular cylinder.
In this framework, the equation governing the proton concentration on the surface can be written as follows,
\begin{eqnarray}
% \begin{aligned}
   \frac{\partial h(s,t)}{\partial t}&=&D\frac{\partial^2 h(s,t)}{\partial s^2}+S_{in}(s,t)-S_{out}(s,t) \\ \label{eq:protonfieldequation} 
%   \end{aligned}
%   \begin{aligned}
     \left.\frac{\partial h(s,t)}{\partial s}\right|_{s=0}&=&0\label{eq:cristaejunction}
\end{eqnarray}
where Eq. (\ref{eq:cristaejunction}) illustrates the reflecting barrier, $D$ is the diffusion coefficient and $S_{in}(s,t)$ and $S_{out}(s,t)$ are the source and the sink of proton respectively.
We consider a system oscillating with a period $T=2\pi/\omega$ between a state IV of  homogeneous proton concentration and a state III associated with a maximal proton flux and write
\begin{eqnarray}
		S_{in}(s,t)&=&\frac{k_{sc}\left(1-\cos(\omega t)\right)}{2\sqrt{2\pi\Delta_1^2}}\exp{\left(\frac{-(s-L_s)^2}{2\Delta_1^2}\right)}\\
		\label{sourceout}
		S_{out}(s,t)&=&\frac{k_{sk}\left(1-\cos(\omega t)\right)}{2\sqrt{2\pi\Delta_2^2}}\left([{\rm H}^+]_{\rm IV}+h(s,t)\right)\nonumber\\ &\times &\exp{\left(\frac{-(s-L)^2}{2\Delta_2^2}\right)}
\label{sourceout}
\end{eqnarray}
with $k_{sc}$ the maximal rate of injection of the proton source, $k_{sk}([{\rm H}^+]_{\rm IV}+h(s,t))$ the maximal rate of the proton sink. The spatial extensions of the source and of the sink are modeled by two Gaussian functions of widths $\Delta_1$ and $\Delta_2$ respectively. The system oscillates between state IV ($S_{in}(s,t)$=0, $S_{out}(s,t)$=0) for $t_{{\rm IV}}=0$ modulo T, noted $[T]$, and state {\rm III} (where the source and the sink function at their top rates) for $t_{{\rm III}}=T/2$ $[T]$ with a frequency $\omega$.
We assume that there is no proton accumulation in the cristae during a period, {\it i.e.} $\int_0^L ds\int_0^{2 \pi/\omega} dt S_{in}(s,t)=\int_0^L ds\int_0^{2 \pi/\omega} dt S_{out}(s,t)$, which sets the value of the ratio $k_{sc}/k_{sk}$.

The proton concentration $h(s,t)$ is obtained by solving numerically Eq. (\ref{eq:protonfieldequation})  with an initial vanishing concentration field $h(s,0)=0$, using the parameter values given in Table I. The profiles of the proton concentration along the tube at different times, shown in Fig. 3, are obtained assuming  a diffusion coefficient of $D$=10$^{-7}$ cm$^2$.s$^{-1}$ \cite{gennis2016}, which corresponds to the estimated diffusion coefficient of protons along a lipid membrane. The system oscillates between state IV, in which the source $S_{in}(s,t)$ and the sink $S_{out}(s,t)$ vanish and the field $h(s,t_{\rm IV})$ is  uniform and equal to zero along the cylinder and state III, for which the proton concentration is approximately homogeneous between the junction ($L$=0) and the source ($L=L_s$) and decreases between the source and the sink. Note that protons diffuse with a characteristic time $\tau_D=L^2/D$=0.25 ms between the source and the sink. An injection rate $k_{sc}$ equal to 600 protons per second at the maximum rate, which is a reasonable value for a crista of this size\cite{rieger2014lateral}, leads to a concentration of  $8\,10^{-2}$ proton per nanometer. These plots are obtained in the case of
an oscillating period much longer than the typical diffusion time along the tube, $2 \pi/ \omega \gg L^2/D$.
 %\color{blue}Therefore the fields in state III and IV are close to the equilibrium solutions obtained at zero and maximal fluxes respectively. We will remain in this regime henceforth. Dire pourquoi c'est le regime realiste? Et aussi, comparer aux temps de relaxation de la forme de la membrane? Autrement dit, est-on quasiment a l'equilibre a chaque concentration?\color{black}

	\begin{table}
		\renewcommand{\arraystretch}{1.8}
\begin{tabular}{|c|c|c|c|}
		\hline
	$L$ & 150 nm &	$R$ & 10 nm \\
	\hline
		$k_{sc}$ &600 s$^{-1}$\cite{rieger2014lateral} &	$k_{sk} $&  2.7.10$^4$ s$^{-1}$ \cite{rieger2014lateral} \\
		\hline
$D$ & 10$^{-7}$ cm$^2$.s$^{-1}$\cite{gennis2016} & $\omega$ & 0.63 s$^{-1}$ \\
\hline
	$\sigma_0$ & 10$^{-7}$ N.m$^{-1}$\cite{AF2} &
		$\kappa_0$ & 10$^{-19}$ N.m\\
		\hline
		$\eta_b$ & 10$^{-9}$ N.s.m$^{-1}$\cite{AF2} &
			$\eta_s$ & 5.10$^{-10}$ N.s.m$^{-1}$\cite{AF2}\\
		\hline
	\end{tabular}
\caption{Parameter values estimated from recent experimental measurements.}
	\end{table}

\begin{figure}
    \includegraphics[width=\columnwidth]{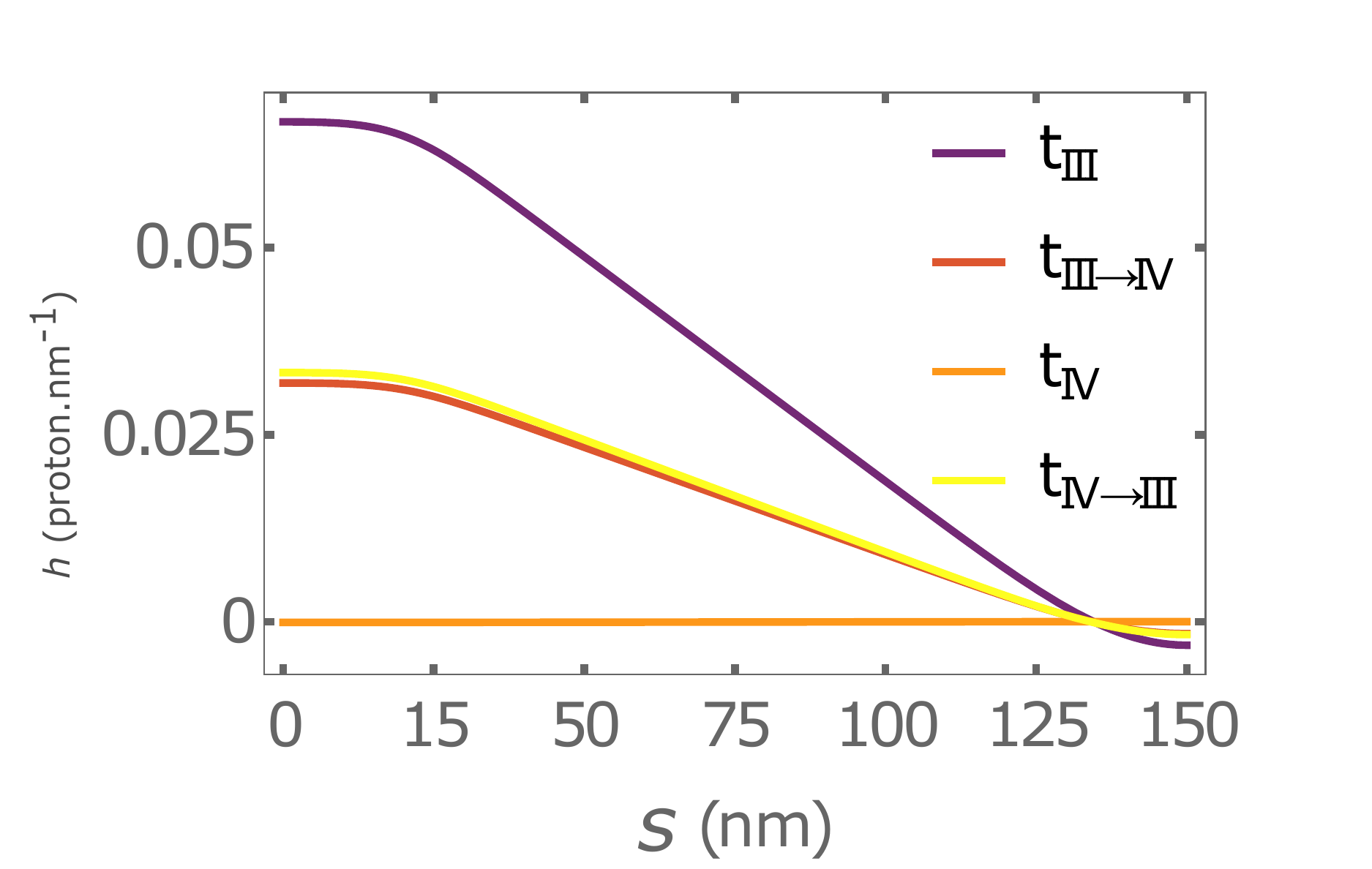}
    \caption{Dynamics of the proton field along the cristae. The plots represent the field $h(s,t)$ solution of Eq. (\ref{eq:protonfieldequation})  for parameter values given in Table I and for $t_{{\rm IV}}=0$ $[T]$, $t_{{\rm IV}\rightarrow {\rm III}}=T/4$  $[T]$,
 $t_{{\rm III}}=T/2$ $[T]$, $t_{{\rm III}\rightarrow{\rm IV}}=3T/4$ $[T]$.
 The concentration $h(s,t)$ is expressed in proton per nm.}
    \label{fig:characteristiclengths}       
\end{figure}

\subsection{Model of the membrane}
To model the membrane of a mitochondrial crista, we start from the Helfrich model  for elastic membranes.
Developed by Wolfgang Helfrich in 1973\cite{helfrich1973}, this effective energy functional of a membrane takes into account molecular properties of lipid membranes such as the fluidity and the absence of in-plane shear stress but is written at a continuous coarse-grained scale.
We modify the standard model by introducing pH dependent parameters, 
\begin{align}
H=\int_{\Omega}[\frac{1}{2}\kappa(s)(C-C_0(s))^2+\sigma(s)]dA\label{eq:hamiltonianhelfrich},
\end{align}
where $C$ is the local curvature of the surface and $dA$ the area of a surface element.
The surface tension $\sigma(s)$, the bending modulus $\kappa(s)$, and the spontaneous curvature parameter $C_{0}(s)$ are assumed to depend linearly on the surface proton concentration as follows,
\begin{eqnarray}
\label{eq:helfrichphenomenologyKappa}
\kappa(s) &=&\kappa_0+h(s)\delta\kappa\\
\label{eq:helfrichphenomenologySigma}
\sigma(s) &=&\sigma_0+h(s)\delta\sigma\\
\label{eq:helfrichphenomenologyC00}
C_0(s) &=&C_{00}+h(s)\delta C_0.
\end{eqnarray}
with $h(s)$ defined in Eq. (\ref{eq:defn_h(s)}).
For simplicity the Gaussian curvature is not considered here.
In this chemico-mechanical model, the chemical reaction of ATP synthesis generates a dynamical field $h(s,t)$ on the membrane and will drive the deformation of the cylinder.

\section{Dynamical equations of the surface deformation}
We consider an initial equilibrium state defined by a vanishing field $h(s,t)=0$ and a finite cylinder of length $L$ and of radius $R$. In this case, the Hamiltonian given in Eq. (\ref{eq:hamiltonianhelfrich}) restricts to the Helfrich model in which the spontaneous curvature $C_{00}$ is a phenomenological parameter, illustrating an asymmetry in the membrane.  
A non-vanishing, non homogeneous proton field $h(s)$ will lead to a deformed cylinder. The initial surface and the deformed surface, can be respectively parametrized by the three-dimensional vectors ${\bf X}_0$ and ${\bf X}_t$,  defined as follows:
\begin{eqnarray}
 {\bf X}_t={\bf X}_0+{\bf \delta X}_t, \end{eqnarray}
with
\begin{eqnarray}
{\bf X}_0&=&\left(\begin{array}{c}
     R\cos(\theta) \\
     R\sin(\theta) \\
     z \end{array}\right), \quad 
{\bf \delta X}_t=\left(\begin{array}{c}
     u_n(z,t)\cos(\theta) \\
     u_n(z,t)\sin(\theta) \\
     u_s(z,t) \end{array}\right) \nonumber \\
     \label{surfaces} 
     &{\rm with}& \quad z \in [0,L], \quad \theta \in [0, 2\pi]. 
\end{eqnarray}
The surface is thus parametrized by the variables $z$, $\theta$ such that any point on this surface can be uniquely represented by a value of each of these parameters: $\mathbf{X}_t(s,\theta)$.
The deformed state ${\bf X}_t$ is characterized by two fields $u_n(z,t)$ and $u_s(z,t)$ defined in Fig. 2B.
In this work, we consider only small deformations, {\it i.e.} $u_s(z,t)/R$, $u_n(z,t)/R$ much smaller than 1.
In the following, we will work in the intrinsic basis of the deformed surface represented in Fig. 2B, and use the curvilinear abscissa $s$. To first order, the derivatives with respect to $z$ and $s$ of the deformation fields are equal: $\partial_z u_i(z,t)=\partial_s u_i(s,t)$, with $i=s,n$, and we will use the second ones.

\subsection{Some elements of differential geometry for an axisymmetric membrane}
We wish to describe the dynamics of the axisymmetric membrane ${\bf X}_t$ driven by the concentration field $h(t)$.   Let us first introduce some terminology and results from differential geometry and their expressions to the first order in the deformation field. 
Note that to simplify the notations, we denote the fields $u_n(s,t)$ and $u_s(s,t)$ as $u_n$ and $u_s$ and the spatial derivative  $\partial_s u_i(s,t)=u'_i$, with $i=s,n$.
%\begin{figure}
%    \centering
%    \includegraphics[scale=0.7]{diff_geom.png}
%    \caption{An arbitrary surface embedded in 3D space and its intrinsic coordinate system}
%    \label{fig:diff_geom}
%\end{figure}
The tangent vectors on the surface are defined as
\begin{eqnarray}
\mathbf{e}_s&=&\partial_{s}\mathbf{X_t}=\begin{pmatrix}
u_n'\cos(\theta) \\ u_n'\sin(\theta)\\1+u_s'(s)
\end{pmatrix},\nonumber\\ \quad \mathbf{e}_\theta&=&\partial_{\theta}\mathbf{X_t}=\begin{pmatrix}
-(R+u_n)\sin(\theta)\\(R+u_n)\cos(\theta)\\0
\end{pmatrix}.
\label{eq:defn_tangents}
\end{eqnarray}

The normal vector can be expressed as
 \begin{equation}
 \mathbf{n}=\frac{e_\theta\wedge e_s}{\mid e_\theta\wedge e_s\mid}=\begin{pmatrix}
\cos(\theta)\\sin(\theta)\\-u_n'
 \end{pmatrix},
 \end{equation}
and the metric of the surface is defined as $g_{ab}={\bf e_a}\cdot{\bf e_b}$, with $(a=(\theta, s),b=(\theta, s))$
and is equal to 
\begin{align}
g_{ab}\approx  \begin{pmatrix}
R^2+2Ru_n & 0\\
0 & 1+2u'_s
\end{pmatrix}.
\end{align}
 
The curvature tensor, also  known as the \textit{second fundamental form}, is defined as $K_{ab}=
\mathbf{e}_a\cdot\partial_{b}\mathbf{n}$ 
using the convention that for a pointing outward normal vector, the curvature is positive  \cite{Deserno_curvatureconvention}. It gives :
\begin{align}
K_{ab}=\begin{pmatrix}
R+u_n &0\\0 & -u_n''
\end{pmatrix}.
\end{align}
Finally, the sum $C=K^s_s+K^{\theta}_{\theta}$ of the principal curvatures can be written as,
\begin{align}
    C=\frac{1}{R}-(\frac{u_n}{R^2}+u_n'')\label{eq:curvaturecylinder}.
\end{align}
using $K^a_b=K_{ak}g^{kb}$, with $g^{ab}=(g_{ab})^{-1}$.

Finally, we recall the expression of the covariant derivative of a tangential vector $x^a{\bf e}_a$, 
\begin{equation}
\nabla_ax^b=\partial_ax^b+\Gamma^b_{ac}x^c \label{eq:defn_covariant_derivative}
\end{equation}
and of a tensor $t^{ab}{\bf e}_a \otimes {\bf e}_b$,
\begin{equation}
\label{covaraintderivativetensor}
\nabla_at^{bc}=\partial_a t^{bc}+\Gamma^b_{ad}t^{dc}+\Gamma^c_{ad}t^{bd},
\end{equation}
where the Christoffel symbols can be written as
\begin{equation}
\label{christofel}
\Gamma_{ab}^s=\left(\begin{array}{cc} -R u_n' &  0 \\ 0& u_s'' \end{array}\right), \quad \Gamma_{ab}^{\theta}=\left(\begin{array}{cc} 0 &\frac{u_n'}{R}  \\  \frac{u_n'}{R} & 0 \end{array}\right).
\end{equation}

To characterize the dynamical deformation of the tube, we introduce 
the flow velocity of the surface elements of the membrane,
\begin{equation}
{\bf v}(s,t)=v_s(s,t){\bf}e_s+v_n(s,t){\bf n} \label{eq:velocity}
\end{equation}
with $v_s=\partial_t u_s$ and $v_n=\partial_t u_n$. It is composed of an in-plane flow ${\bf v}_s=v_s{\bf e}_s$ and a term describing the deformation of the surface ${\bf v}_n=v_n{\bf n}$.

\subsection{Stress tensor acting on the deforming surface}
%The deformation velocity of the membrane
%\begin{align}
%{\bf v}(s,t)=v_z(s,t){\bf}e_z+v_n(s,t){\bf n}
%\end{align}
%To determine the dynamical equations of the deformation of the cristae driven by an out-of-equilibrium concentration field $h(s,t)$, we first apply the Stokes approximation,{\it i.e} we neglect all inertial effects and simply requires that all forces balance.

Next, we determine the stress tensor {\bf f} acting on the surface ${\bf X}_t$. 
General surface stresses are complex objects that one can grasp by asking the question: ``What forces should be exerted onto a membrane edge with unitary length to prevent it from shrinking?'' \cite{Deserno_curvatureconvention}. In this case, the stress tensor 
is the sum of two contributions: a mechanical stress tensor ${\bf f}_{H}$ deriving from the Helfrich energy given in Eq. (\ref{eq:hamiltonianhelfrich}) and a viscous stress tensor ${\bf f}_{\eta}$. 

The mechanical stress tensor ${\bf f}_H$ can be written as a 3x2 tensor\cite{Capovilla02,fournier2007} that can be decomposed into a surface stress tensor $f^{ab}_H$ generating forces tangent to the surface and a 2x1 tensor ${\bf f}^{n}_{H}=f^{an}_H{\bf n}\times{\bf e}_a$ generating forces normal to the surface. To derive the expression of {\bf f},
% deriving from Hamiltonian given in Eq. (\ref{eq:helfrichphenomenology}q:helfrichphenomenology}) associated with non constant parameters, 
we follow the approach developed by Guven and coworkers, presented in Appendix A.
%, which involves a minimization under constrain of the Hamiltonian yielding the surface energy. 
Instead of varying the shape of the membrane ( ${\bf X}_t\rightarrow {\bf X}_t+\delta {\bf X}_t $) and explicitly tracking the changes of the intrinsic basis, the metric, the curvature and the energy of 
the deformed membrane, this elegant approach enforces the geometric relations  associated with the fundamental forms of the membrane by introducing Lagrange multipliers.
We introduce the extended functional
\begin{widetext}
\begin{eqnarray}
\label{Hc}
H_c&=&H+\int_{\Omega}\lambda^{ab}(g_{ab}-\mathbf{e}_a\cdot\mathbf{e}_b)dA+\int_{\Omega}\Lambda^{ab}(K_{ab}-\mathbf{e}_a\cdot\nabla_b\mathbf{n})dA+\int_{\Omega}\mathbf{f}^a\cdot(\mathbf{e}_a-\nabla_a\mathbf{X})dA \nonumber\\
&+&\int_{\Omega}\lambda^a_{\bot}(\mathbf{e}_a\cdot\mathbf{n})dA+\int_{\Omega}\lambda_n(\mathbf{n}^2-1)dA.
\label{eq:fullhamiltonianwithlagrangemultipliers}
\end{eqnarray}
\end{widetext}
in  which $H$ is given in Eq. (\ref{eq:hamiltonianhelfrich}), the matrices $\lambda^{ab}$ and $\Lambda^{ab}$ enforce the definition of metric and the curvature, $\mathbf{f}^a$ pins the basis vector to the tangent of the surface, $\lambda^a_{\bot}$ enforces the normal vector to be perpendicular and $\lambda_n$ its normalisation.
The minimisation of the Hamiltonian Eq. (\ref{Hc}) with respect to the, now, 12 independent functions 
 gives\cite{Muller_phd}
\begin{eqnarray}
\label{fabH}
\mathbf{f}_H^{ab}&=T^{ab}-\mathcal{H}^{ac}K^b_c \\ \label{fanH}
f_H^{an}&=-(\nabla_b\mathcal{H}^{ab})
\end{eqnarray}
with
\begin{eqnarray}
	\mathcal{H}^{ab}=\frac{\delta\mathcal{H}}{\delta K_{ab}}\\
	T^{ab}=-\frac{2}{\sqrt{g}}\frac{\delta \sqrt{g}\mathcal{H}}{\delta g_{ab}}
	\label{eq:expressionsforTabHab}
\end{eqnarray}
with ${\mathcal H}=\frac{1}{2}\kappa(s)(C-C_0(s))^2+\sigma(s)$, the mechanical energy density of the membrane
(Details of the calculation are given in Appendix A).

In our case, the mechanical stress associated to the surface ${\bf X}_t$ depends both on the deformation fields $(u_s(s,t),u_n(s,t))$ and on the concentration field $h(s,t)$. The surface stress tensor $f^{ab}_H$ and the normal stress tensor $f^{an}_H$ expanded to first order in the fields can be expressed as 
\begin{eqnarray}
f_H^{ab}&=&f_{H0}^{ab}+f_{H1}^{ab} \label{eq:splittingtensorterms} \\
f_H^{an}&=&f_{H0}^{an}+f_{H1}^{an}
\end{eqnarray}
with 
\begin{eqnarray}
\label{f0}
f_{H0}^{ab}&=&\left(\begin{array}{cc} \frac{\kappa_0(1-X^2)-2\sigma_0R^2}{2R^4}  & 0\\
0 &    -\frac{(1-X)^2\kappa_0}{2R^2}-\sigma_0 \end{array}\right),
\end{eqnarray}
and $f^{an}_{H0}=0$, and where the first order part,
\begin{eqnarray}
\label{fabH}
f^{ab}_{H1}&=&f^{ab}_{h}+f^{ab}_{M}\\
\label{fanH}
f^{an}_{H1}&=&f^{an}_{h}+f^{an}_{M}
\end{eqnarray}
is the sum of a term $f^{ab}_{1h}$ depending on the concentration field $h(s,t)$ and of a term $f^{ab}_{1M}$ depending on the deformation fields $u_n(s,t),u_s(s,t)$.

Using the expression of the Hamiltonian given in Eq. (\ref{eq:hamiltonianhelfrich}), the dependences of $\kappa$, $C_{00}$ and $\sigma$ on $h(s,t)$ given in Eqs. (\ref{eq:helfrichphenomenologyKappa})-(\ref{eq:helfrichphenomenologyC00}) and the expression of the stress tensor given in Eq. (\ref{fabH}, \ref{fanH}), we derive the expression of the stress tensor to first order in the fields and find,
\begin{widetext}
	\begin{eqnarray}
	\label{fabm}
	f^{ab}_{1M}&=&\left(\begin{array}{cc} \frac{(2R^2\sigma_0+(2-X^2)\kappa_0)u_n(s)-R^2X\kappa_0u_n''(s)}{R^5}  & 0\\
	0 & \frac{1-X}{R^3}\kappa_0u_n(s)+\frac{(1-X)^2\kappa_0+2R^2\sigma_0}{R^2}u_s'(s)\end{array}\right)  \\
	f^{an}_{1M}&=&\left(0, \frac{\kappa_0}{R^2}u'_n(s)+\kappa_0u'''_n(s) \right)
	\end{eqnarray}
	\begin{eqnarray}
	\label{fabh}
	f^{ab}_{1h}&=&\left(\begin{array}{cc}
	 -\frac{2R^2\delta\sigma+2RX\kappa_0\delta C_0-(1-X^2)\delta\kappa}{2R^5}Rh(s) & 0 \\
	0 & \frac{2R((1-X)\kappa_0\delta C_0-R\delta\sigma)-(1-X)^2\delta\kappa}{2R^2}h(s) \end{array}\right)\\
	f^{an}_{1h}&=&\left(0, \frac{R\kappa_0\delta C_0-(1-X)\delta\kappa}{R}h'(s)\right) .
	\end{eqnarray}
\end{widetext}

The stress tensor depending on the proton field  involves both the field $h(s,t)$ in its tangential component  $f^{ab}_{1h}$, and its spatial derivative $h'(s,t)$ in its normal component   $f^{an}_{1h}$. The latter contribution will vanish for a constant field $h$.
In fact, a constant field $h$ simply induces a renormalisation of the parameters of the Helfrich model.
By contrast, a spatially non homogeneous field $h(s,t)$ will generate forces on the normal direction that tend to pinch  or expand the tube.

The viscous stress tensor for a deforming membrane is a surface tensor that can be obtained from the two-dimensional strain rate,
\begin{align}
	v_{ab}=\frac{1}{2}\left(\nabla_a v_b +\nabla_b v_a\right)+K_{ab}v_n
\end{align} 
where ${\bf v}$ is given in Eq. (\ref{eq:velocity}), that is derived from the strain tensor of a three dimensional fluid shell taken in the limit of a small thickness\cite{thesemietke}. 
The viscous stress in compressible thin films can be expressed as
\begin{equation}
\label{viscousstress} 
f_{ab,\eta}=2\eta_s(v_{ab}-\frac{1}{2}v^c_cg_{ab})+\eta_bv^c_cg_{ab},
\end{equation}
which involves the two-dimensional strain rate $v_{ab}$ and phenomenological coefficients $\eta_s$ and $\eta_b$ that are the shear and bulk viscosity of the film. The viscous stress  ${\bf f}_{\eta}=f_{ab,\eta}{\bf e}^a\times {\bf e}^b$ can be written as 2x2 matrix, whose components are given in covariant coordinates in Eq. (\ref{viscousstress}) \cite{Deserno_curvatureconvention}. It yields the force generated by the flow within the membrane surface. Note that we do not account for the bulk viscosity of the fluid surrounding the membrane in our force balance. Indeed, its contribution can be neglected for strongly curved membrane buds~\cite{Arroyo09} and tubes~\cite{Rahimi12} with characteristic sizes smaller than a few micrometers, which is the appropriate regime for cristae.

In the absence of deformation, i.e. when the fields $v_n$, $v_s$, $u_n$, $u_s$, $h$ vanish, the viscous stress also vanishes and the equilibrium shape of the surface can be obtained by setting to zero the divergence of the stress tensor $f^{ab}_{H,0}$ given in Eq. (\ref{f0}).

\subsection{Hydrodynamic equations for the tubular membrane}
We now consider the response of the membrane to a time varying field, $h(s,t)$.
In the absence of inertia, corresponding to the low-Reynolds number regime which is appropriate at the lengthscales considered, the dynamics of the system derives from the force balance, which can be written in the tangential basis of the surface as,
\begin{eqnarray}
\nabla_af^a_b+K_{ab}f^a_n&=&0\label{eq:forcebalance2},\\
\nabla_af^{an}-K_{ab}f^{ab}&=&0 \label{eq:forcebalance1}.
\end{eqnarray}
where the stress tensor ${\bf f}={\bf f}_{H}+{\bf f}_{\eta}$ is the sum of the viscous and the mechanical stress tensor. The first equation,  Eq. (\ref{eq:forcebalance2}), corresponds to the force balance on the surface along ${\bf e}_s$ and ${\bf e}_{\theta}$. The third equation, Eq. (\ref{eq:forcebalance1}), often called the shape equation of the surface is the force balance in the normal direction.
%The two equations above can also be recognized to be the force balance equations on a membrane. Thus the equation for the shape being in equilibrium is equivalent to a balance of forces, which is quite intuitive.
Expanding the covariant derivatives in Eqs. (\ref{eq:forcebalance2} ,\ref{eq:forcebalance1}) and collecting the first order terms in the fields $v_n$, $v_s$, $u_n$, $u_s$, $h$,  
%\begin{eqnarray}
%&\partial_s& f^s_{1s}+\left(f_{0H s}^s-f_{0\theta}^{\theta}\right)\Gamma^s_{s \theta}%+Rf^{ns}=0,\\
%&\partial_s& f_{H1}^{ns}-u_s f_1^{ss}-Rf_{0H}^{ss} +u_n''f_1^{\theta \theta}=0.
%\end{eqnarray}
%with $f^a_{b1}=f^a_{\eta,b}+f^a_{H1,b}$. The equation along the direction ${\bf e}_{\theta}$ vanishes for symmetry reasons.
we derive the hydrodynamic equations governing the time evolution of the system
\begin{widetext}
	\begin{eqnarray}
	\label{eq:forcebalance1withvelocites}
	(\eta_s+\eta_b)\partial_sv^s_s+(\eta_b-\eta_s)\partial_sv^{\theta}_{\theta}+\partial_s f^s_{H1,s}+\left(f_{H0,s}^s-f_{H0,\theta}^{\theta}\right)\Gamma^s_{s \theta}+Rf^{ns}&=&0 \\ 
	(\eta_s+\eta_b)\frac{v^s_s}{R}+(\eta_b-\eta_s)\frac{v^{\theta}_{\theta}}{R}+\partial_s f_{H1}^{ns}-u_s f_{H0}^{ss}-Rf_{H1}^{ss} +u_n''f_{H0}^{\theta \theta}&=&0.
	\label{eq:forcebalance2withvelocites}
	\end{eqnarray}
\end{widetext}
The force balance along ${\bf e}_{\theta}$ vanishes for symmetry reasons.
Replacing the mechanical tensors $ f^{ab}_{0H}$, $f^{ab}_{1H}$, $f^{an}_{1H}$ by their expressions given in Eqs. (\ref{f0},\ref{fabH},\ref{fanH}), we finally obtain the dynamical equations of the system as a function of the velocity, displacement and concentration fields: 
\begin{eqnarray}
\label{forcetangentcoef} 
&a_1&v_n'+a_2v_s''+a_3h'=0\\
& b_1&v_n+b_2v'_s+b_3u_n+b_4u_n''+b_5u_n''''\nonumber\\ &+&b_6h+b_7h''=0\label{forcenormcoeff}
\end{eqnarray}
where the coefficients $a_i$, (i=1,...,7) and $b_i$, (i=1,...,3) are given in Appendix \ref{app:coeffsdyn} and where $u_i'$ and $v'_i$ note the spatial derivative $u_i'=\partial_s u_i(s,t)$ and $v_i'=\partial_s v_i(s,t)$  for $i=s,n$.

\section{Static Green function of an infinite membrane cylinder}
To gain physical insight on the model and on the static solutions of Eqs. (\ref{eq:forcebalance1withvelocites}, \ref{eq:forcebalance2withvelocites}), we derive the static Green function of the system.
We consider as a reference equilibrium state an infinite cylinder of radius $R$ and a vanishing field $h(s)$=0. 
The shape equation for this system, given in Eq. (\ref{eq:forcebalance1}), can be written as
\begin{equation}
	\sigma_0-\frac{(1-X^2)\kappa_0}{2R^2}=0,
\end{equation}
where we have employed the expression of $f^{ab}_{H0}$ given in Eq. (\ref{f0}).
Given that $R$, $\sigma$ and $\kappa$ are necessarily positive, this equation admits a solution for 
\begin{equation}
\label{limX}
X=C_{00}R \quad \in  \quad ]-1,1[.
\end{equation}
In this range (see Fig. 4A), the radius of the equilibrium cylinder $R$ can be expressed as a function of the parameters of the non-perturbed Helfrich model,
\begin{equation}
R=\frac{1}{\sqrt{C_{00}^2+\frac{2\sigma_0}{\kappa_0}}}.
\label{eq:relationbetweenparameters}
\end{equation}
The stationary shape of a deformed cylinder associated with a perturbation field $h(s)\neq 0$ is given by the displacement fields $(u_n(s), u_s(s))$ that are solutions of the Eqs. (\ref{forcetangentcoef},\ref{forcenormcoeff}) where $v_n$ and $v_s$ are set to zero. 
Replacing the coefficients ($a_i$, $b_j$) by their expressions given in Appendix \ref{app:coeffsdyn}, we obtain,
\begin{widetext}
	\begin{eqnarray}
	\left(\frac{4R\sigma_0\delta C_0}{1+X}-2\delta\sigma-\frac{\delta\kappa(1-X)^2}{R^2}\right) h'(s)&=&0\\
	\label{eq:eq22}
	-\frac{2\sigma_0}{R^2(1-X^2)}\left(u_n(s)+2R^2Xu''_n(s)+R^4u''''_n(s)\right)&=&\frac{2((1+X)\delta\sigma-R\sigma_0\delta C_0)}{R(1-X^2)}(h(s)+R^2h''(s)).
	\label{eq:shapehelfrichexpressionmech}
	\end{eqnarray}
\end{widetext}

The first equation corresponds to the force balance along the vector ${\bf e}_s$. The second equation is the shape equation. We observe that this system of equations does not depend on $u_s$. A non-homogeneous perturbation field $h(s)$, which yields non-homogeneous tension, bending rigidity and spontaneous curvature, can lead to a stationary shape in the cylinder geometry only if the condition 
\begin{equation}
\left(\frac{4R\sigma_0\delta C_0}{1+X}-2\delta\sigma-\frac{\delta\kappa(1-X)^2}{R^2}\right)=0\label{eq:relationbetweenperturbations}
\end{equation} 
between the perturbation parameters, ($\delta \kappa$, $\delta \sigma$, $\delta C_0$), is satisfied. Assuming that this condition holds, we express $ \delta \kappa$ as a function of $\delta \sigma$ and $\delta X$,
\begin{equation}
\delta \kappa= \frac{4R\sigma_0}{\left(1-X\right)^2\left(1+X \right)}\delta C_0-\frac{2R^2}{(1-X)^2}\delta \sigma.
\end{equation}

Let us now consider a localized perturbation in the proton concentration: $h(s)=h_0\delta(s)$ with $h_0=1$ introduced to dimension $h(s)$ , and derive the Green function $G_n(s)$ yielding the deformation field $u_n(s)$ in response to this perturbation. Performing a Fourier transform of the shape equation given in Eq. (\ref{eq:shapehelfrichexpressionmech}) and introducing $\tilde{u}_n(q)=1/2\pi\int ds e^{iqs}u_n(s)$, the normal deformation field in the Fourier space, we find 
\begin{equation}
\tilde{u}_n(q)\equiv \tilde{G}_n(q)=\frac{R(q^2R^2-1)[(1+X)\delta\sigma-R\sigma_0\delta C_0]}{\sigma_0\sqrt{2\pi}(1-2Xq^2R^2+q^4R^4)}.\label{eq:fouriertransformofun}
\end{equation}

Performing the inverse Fourier transform of $\tilde{G}_n$ given in Eq. (\ref{eq:fouriertransformofun}), we obtain the expression of the Green function in real space
%\begin{align}
%   \begin{aligned}
%       G_n(s)=\frac{iA}{q_1q_2}&\sqrt{\frac{\pi}{2}}[q_2(q_1^2R^2-1)e^{-iq_1|s|}\\&+q_1(q_2^2R^2-1)e^{iq_2|s|}],
%  \end{aligned}\label{eq:genericgreensresult}
%\end{align}
%which on inserting the values for the co-factors and roots gives us the resulting response
\begin{equation}
G_n(s)=G_{n} R \sin{(\frac{|s|}{R}\sqrt{\frac{1+X}{2}})}\exp{(-\frac{|s|}{R}\sqrt{\frac{1-X}{2}})}\label{eq:expressionforun}
\end{equation}
where we have introduced 
\begin{eqnarray}
G_{n}&=&G_{n\sigma}(X)\frac{\delta \sigma}{\sigma}+G_{nC}(X)R \delta C_0 \nonumber
\end{eqnarray}
with
\begin{eqnarray}
G_{n\sigma}(X)&=&-\frac{(1+X)}{\sqrt{2}\sqrt{1+X}}, \quad
G_{nC}(X)=\frac{1}{\sqrt{2}\sqrt{1+X}}.
\end{eqnarray}
The deformation induced by a localized perturbation is thus an oscillating and exponentially decaying function, as shown in Fig. 4B1.
The characteristic lengths of oscillation $\lambda_o$ and decay $\lambda_d$ are given respectively by
\begin{eqnarray}
\lambda_o&=R\sqrt{\frac{2}{1+X}}\label{eq:oscillationlength}\\
\lambda_d&=R\sqrt{\frac{2}{1-X}}\label{eq:decaylength}
\end{eqnarray}
for $X$ satisfying the condition given in Eq. (\ref{limX}).

Fig. 4B2 represents $\lambda_o$ and $\lambda_d$ as functions of $X$ for the range of possible equilibrium cylinders. The decay length $\lambda_d$ is an increasing function of $X$ that diverges for $X=1$. Thus, cylinders with larger spontaneous curvatures are deformed on a longer range by a heterogeneous $h$. The oscillation length $\lambda_o$, on the contrary, is a decreasing function of $X$.

The limit $X=1$  is associated with a phenomenon of buckling ($\lambda_d=0$, $\lambda_o=R$) {\it i. e.} an infinite oscillating deformation wave.  Indeed, $X=1$ corresponds to a cylinder with curvature along $e_{\phi}$ equal to the spontaneous curvature of the membrane. 
A sphere of radius $R$ is an equilibrium shape of such a membrane. A perturbation of the cylinder will tend toward such shapes by forming a succession of drops. Conversely, the limit $X=-$1 corresponds to a cylinder folded in the direction opposite to that of the spontaneous curvature. The deformation induced by a perturbation will then be purely decaying with a characteristic length $\lambda_d=$R, thus minimizing the energy of deformation.

The respective amplitudes $A_{n \sigma}$, $A_{n C}$ of the deformations induced by perturbations in $\sigma$ or in $C_0$, and defined as 
\begin{eqnarray}
A_{n \sigma}(X)&=&G_{n\sigma}(X) \frac{G_n(s_m)}{G_n},\\
A_{n C}(X)&=&G_{nC} (X)\frac{G_n(s_m)}{G_n},
\end{eqnarray}
are plotted with respect to $X$ in Fig. 4B3. Here, we have introduced $s_m=\sqrt{2}/\sqrt{1+X}\sin^{-1}(\sqrt{X+1}/\sqrt{2})$, which is the coordinate of the maximum of $G_n(s)/G_n$.  The absolute values of the amplitudes $A_{n \sigma}$, $A_{n C}$ are increasing functions of X but possess opposite signs.  An increase of the tension ($\delta \sigma >$0, $h(s)>0$) leads to a constriction of the cylinder (since $A_{n\sigma}<0$). On the contrary, an increase of the spontaneous curvature ($\delta C_0>0$, $h(s)>0$) leads to a dilatation of the cylinder (since $A_{nC}>0$). A perturbation in the spontaneous curvature $\delta C_0R=0.1$ (respectively in the surface tension $\delta \sigma/\sigma=0.1$) will induce a variation of the radius of 5 percent (respectively 10 percent), for $X \rightarrow 1$. For $X$ negative or $0< X \ll$ 1, the deformation induced by a variation of $h$ is negligible. 

Fig. 4C shows three-dimensional representations of the non-perturbed  and of the perturbed cylinder for $X=0.9$, and with  $\delta \sigma >0$ and $\delta \sigma <0$, respectively. This illustrates that the model proposed in Eq. (\ref{eq:hamiltonianhelfrich}) can generate tubular membrane shapes of various curvature that can resemble  mitochondrial cristae.

\begin{figure}[h]
	\includegraphics[width=0.5\textwidth]{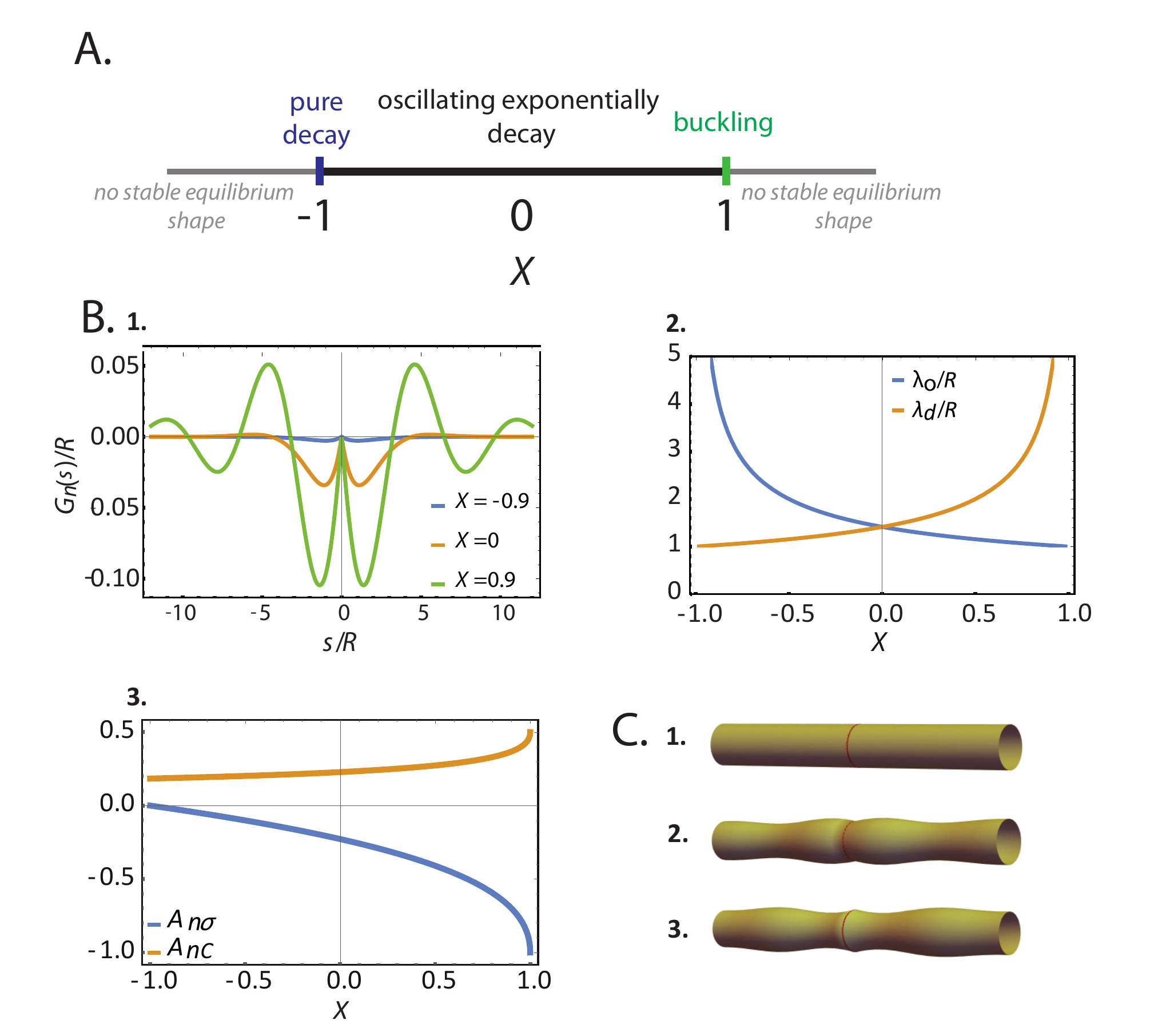}
	\caption{A. Stability diagram of a Helfrich cylinder as a function of $X=C_{00}R$.  B. 1. $G_n(s)/R$, Normalized Green functions for different values of $X=-0.9, 0, 0.9$ and for $\delta \sigma=0.15$ and $\delta C_0=0$. 2. Characteristic lengths of oscillation $\lambda_0$ and decay $\lambda_d$ as a function of  $X$. 3. Maximal amplitude of the deformations driven respectively by a variation of $\sigma$,  $A_{n\sigma}$, and by a variation of $C_{0}$,  $A_{nC}$. C. Three dimensional representations of a cylinder (top) deformed by $h(s)=\delta(s)$ for (X=0.96, $\delta \sigma=0.2$, $\delta \kappa=0$, $\delta C_0=0$)  (middle) and for  (X=0.96, $\delta \sigma=-0.2$, $\delta \kappa=0$, $\delta C_0=0$) (bottom). All quantities are dimensionless. }
\end{figure}

Using the expression of the Green function given in Eq. (\ref{eq:expressionforun}), we can find the stationary shapes generated by any proton concentration field $h(s)$ through
\begin{equation}
u_n(s)=\int_{-\infty}^{\infty}h(x)G_n(s-x)dx.\label{eq:greenstopractical}
\end{equation}
% We can perform numerical integration of \cref{eq:greenstopractical} with a variety of functions to obtain physical deformations. 
As a practical illustration, in Fig. 5,  we consider a step function for $h(s)$, 
\begin{equation}
h(s)=-1 \quad \rm{s<0}, \quad =1 \quad\rm{s>0}
\end{equation}
(see Fig. 5A). The corresponding deformation field $u_n(s)$ obtained using Eq. (\ref{eq:greenstopractical}) is continuous, unlike the input step function, and is an  odd function of $s$ that oscillates before reaching a plateau of constant value, as shown in Fig. 5B. The resulting stationary shape of the tube, represented in Fig 5C, corresponds to two cylinders of different radii welded together through an oscillating neck. 

\begin{figure}
	\includegraphics[scale=0.18]{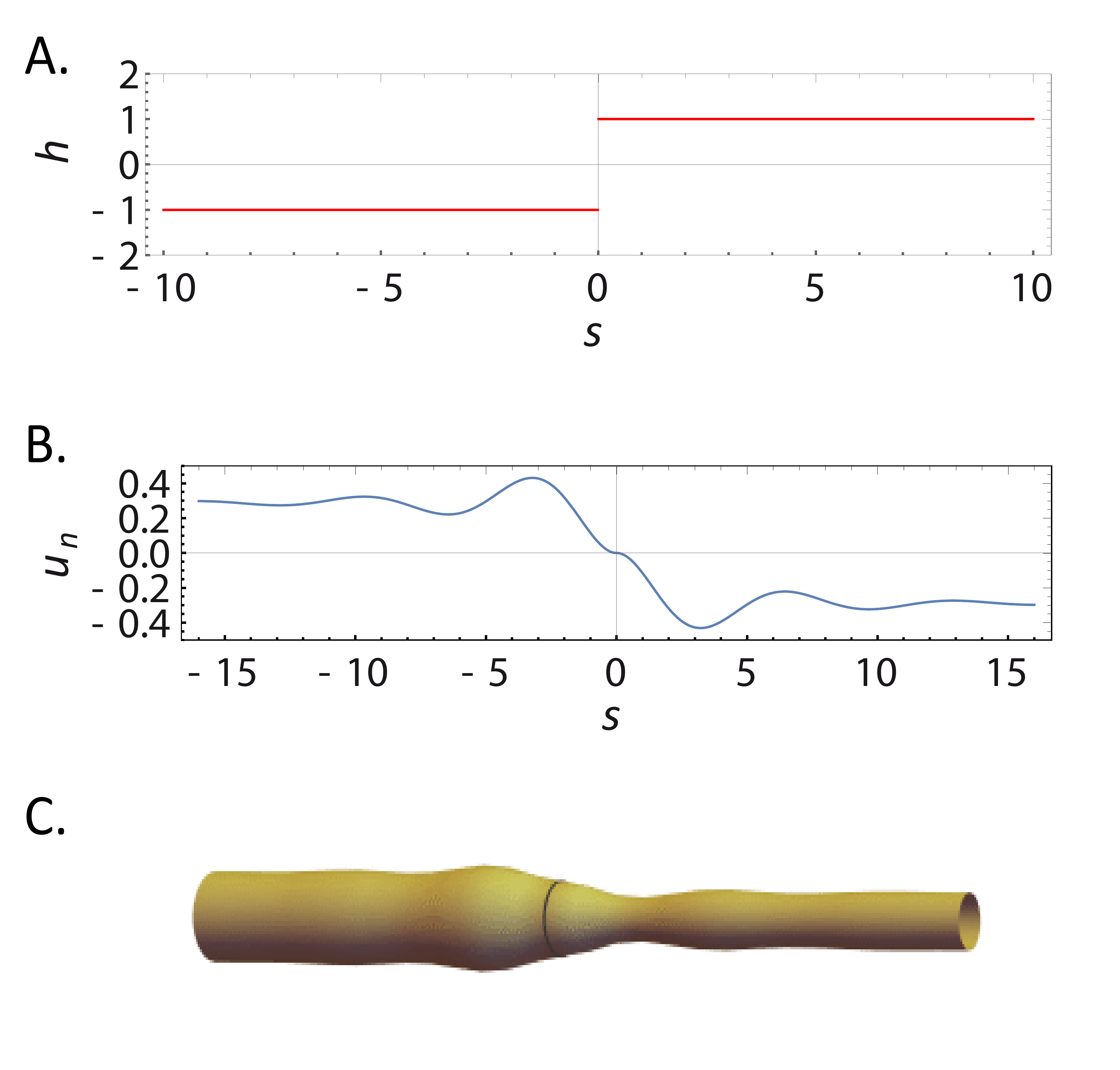}
	\caption{Infinite cylinder submitted to a step of proton concentration. A. Proton concentration $h(s)$ along the cylinder. B. Deformation field $u_n(s)$ induced by a step of proton for $X=0.9$, $\delta \sigma=0.1$, $\delta C_0=-0.1$. C. Three-dimensional representation of the corresponding shape.  All quantities are dimensionless.}
\end{figure}

\section{Application to mitochondria cristae}
In this section, we describe the dynamical deformation of a finite tube of membrane of length $L$ submitted to the proton field $h(s,t)$ solution of Eq. (\ref{eq:protonfieldequation}) and represented in Fig. 3, which models a crista oscillating between state IV and state III. 

To do so, we first specify the external mechanical forces exerted on  the axisymmetric membrane. Here, we do not take into account extra external pressures or viscous forces that could be applied to the membrane and we consider that a constant tension $f_{ext}$ is exerted by the rest of the mitochondria on the tubule boundary rings in $s=0$ and in $s=L$. This force, 
\begin{eqnarray}
f_{ext}(0)&=&-\left( \sigma +(1-X^2)\frac{\kappa_0}{2 R^2}\right) {\bf e}_s, \nonumber\\ f_{ext}(L)&=& \left( \sigma +(1-X^2)\frac{\kappa_0}{2 R^2}\right) {\bf e}_s 
\end{eqnarray}
balances the effective tension of the undeformed Helfrich cylinder, defined by ${\bf X}_0$ in Eq. (\ref{surfaces}), and derives from the zeroth order stress tensor $f_{H0}$ given in Eq. (\ref{f0}). 
Consequently, the first order forces deriving from the cylinder deformation  vanish at the boundary of the tubule, and the first order stress tensors satisfy:
\begin{eqnarray}
\label{forcees}
f^s_{1,s}(0,t)&=&0, \quad  f^s_{1,s}(L,t)=0\\
\label{forcen}
f^{sn}_{H1}(0,t)&=&0, \quad  f^{sn}_{H1}(L,t)=0.
\end{eqnarray}
where $f^s_{1,s}=f^s_{H1,s}+f^s_{\eta,s}$,
and where the expression of the stress is given in Eqs. (\ref{fabH}, \ref{fanH}, \ref{viscousstress}).
Moreover, the edges of the cylinder are assumed to be pinned in $s=0$ and $s=L$, which leads to a vanishing tangential velocity $v_s$ in $s=0$ and $s=L$,
\begin{eqnarray}
 \label{pin}
  v_{s}(0,t)=0, \quad  v_{s}(L,t)=0.
 \end{eqnarray}
In order to facilitate the numerical resolution of the hydrodynamical equations given in Eqs.(\ref{forcetangentcoef}, \ref{forcenormcoeff}), the deformation fields $u_n(s,t), u_s(s,t)$ are expressed as function of the velocity fields $v_n(s,t)$ and $v_s(s,t)$ and of the deformation field at the time $t-dt$, using the backward Euler method and a discretisation of the time with a time step $dt$, 
\begin{eqnarray}
\label{backwardeulers}
u_s^t&=&u_s^{t-dt}+dt\times v_s^t\\
\label{backwardeulern}
u_n^t &=&u_n^{t-dt}+dt\times v_n^t. 
\end{eqnarray}
where we have introduced the notation $(u_i^t, v_i^t)$ for $(u_i(s,t),u_i(s,t)) $.
Inserting the expressions in Eqs. (\ref{backwardeulers},\ref{backwardeulern}) the deformation fields into Eqs. ( \ref{forcetangentcoef}, \ref{forcenormcoeff}) leads to the following coupled system of equations for $v^t_s$ and $v_n^t$
\begin{eqnarray}
\label{eq:forcebalancesnumerics1}
&a_1& \partial_s v^{t}_n+a_2 \partial_s^2 v^{t}_s+a_3 \partial_s h^{t}=0\\
\label{eq:forcebalancesnumerics2}
 &(b_1&+b_3\times dt)v^t_n+b_4\times dt \partial_s v_n^t +b_5\times dt\partial^4_s v_n\nonumber\\ &+&b_2\partial_s v^t_s+b_3 u_n^{t-dt}+b_4\partial_s^2 u^{t-dt}_n\nonumber\\ &+&b_5 \partial_s u^{t-dt}_n+a_6h^t+a_7\partial^2_s h^t=0.
\end{eqnarray}

Assuming that the geometry of the system at the time $t-dt$ is known, the system to solve is a couple of ordinary differential equations in space for $v^t_s$ and $v_n^t$ 
for which the 6 necessary boundary conditions are given in Eqs. (\ref{forcees}-\ref{pin}).
The complete dynamic of the system is obtained by solving Eqs. (\ref{eq:forcebalancesnumerics1}, \ref{eq:forcebalancesnumerics2}) starting at $t=0$, when all the fields, $h^0, u^0_i, v_i^0$, vanish. Then,  knowing the geometry of the system at the time $t-dt$, {\it i. e.}, the deformation fields $u_i^{t-dt}$, and the concentration field at the time $t$, $h^t$,  the velocities $(v_s^{t}(s),v^{t}_n(s))$ are determined as solutions of Eqs. (\ref{eq:forcebalancesnumerics1}, \ref{eq:forcebalancesnumerics2}). Next, the deformation fields $u_n^t$, $u_s^t$ at time $t$ are calculated using Eqs. (\ref{backwardeulers}, \ref{backwardeulern}). The procedure can then be iterated to obtain the state of the system at time $t+dt$, and so forth. 
Further details on the numerical resolution are given in Appendix 3.
\begin{figure}
	\includegraphics[scale=0.45]{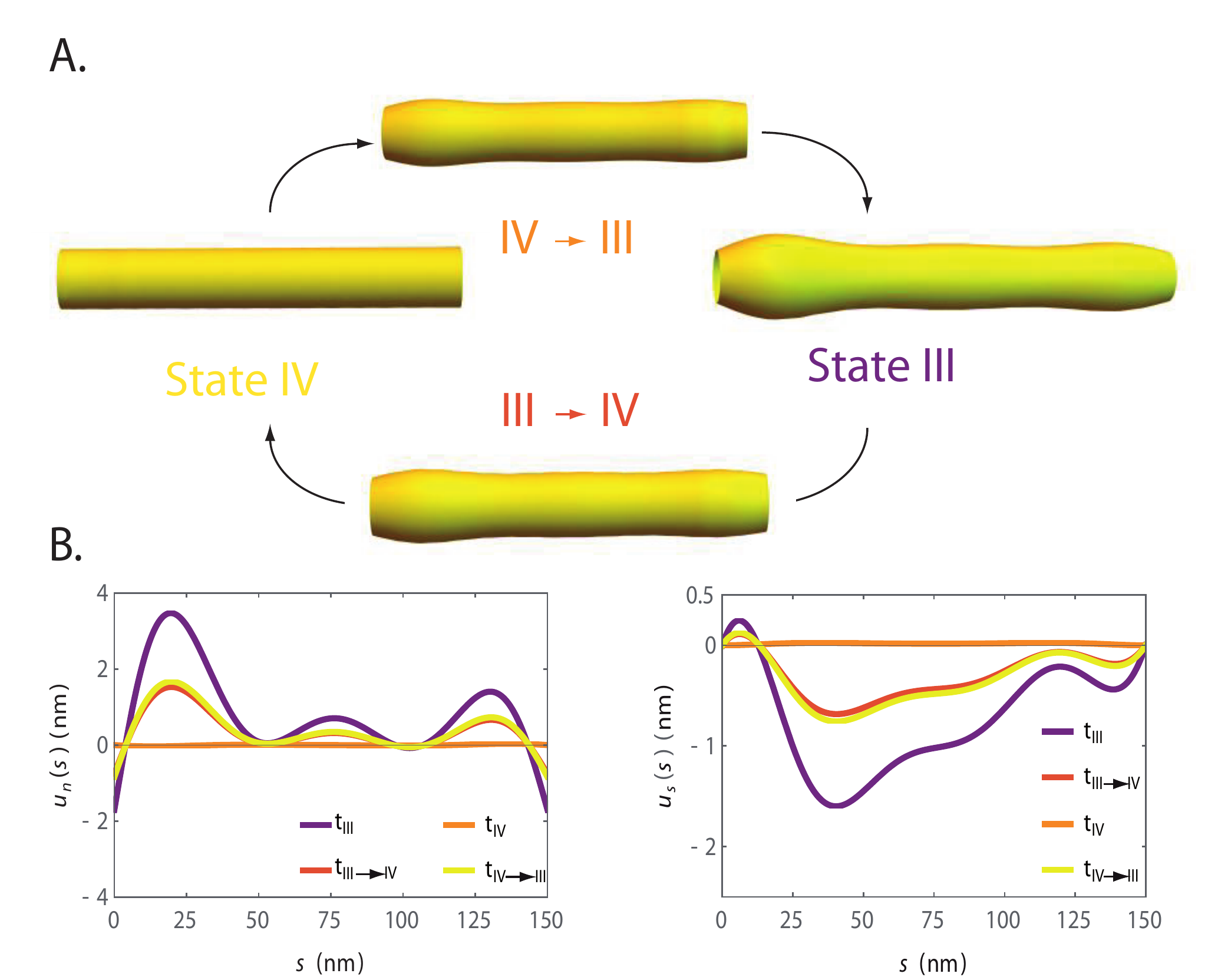}
	\caption{Model crista oscillating between states IV and III. A finite membrane tube is submitted to the proton field $h(s,t)$ represented in Fig. 3.A. Representation of the shape of the cylinder at different times of the period (the same ones as in Fig. 3). B. Deformation fields $u_n(s,t)$ and $u_s(s,t)$  associated to these shapes (same colors as in Fig. 3). The plots are obtained for values given in Table 1, $X=0.8$, $\delta \sigma=-0.35\sigma$ per $h(s,t)$, $\delta C_0=1.05$ per $h(s,t)$.  }
\end{figure}

Fig. 6A represents the shape of the tube for the oscillating proton field given in Fig. 3.  The tube alternates between a quasi-non-deformed cylinder for a quasi vanishing proton field (see Fig. 3) and a tubular invagination presenting a bump and a neck for the top proton flux. These oscillations are obtained in the regime of a fast mechanical relaxation time $\tau_0=2\eta_s\eta_b/\sigma_0(\eta_s+\eta_b)$ compared to the  oscillation period between state III and IV, $T=2\pi/\omega$. It is thus the dynamics of the diffusive process that governs the shape change dynamics in this case. Fig. 6B represents the deformation fields associated with these shapes.  We compute the bump-to-neck ratio obtained in this framework and obtain 1.6. Measuring the projected sizes of the bumps and necks in state III on the 8 best visible cristae on Fig. 1 yields an average bump-to-neck ratio of 1.9 (standard error of the mean: 0.3). The theoretical prediction is in agreement with the measures. This shows that the model developed in this work captures the salient features of the shape of the cristae in state III and IV observed experimentally and represented in Fig. 1 (regular tubes in state IV, bumpy tubes in state III). 

\section{Conclusion}
Mitochondrial cristae are membrane protrusions that confine the ATP-synthesis machinery. Experimental observations have shown that the shape of these protrusions is coupled to the energetic state of mitochondria, as seen in Fig. 1. However, the underlying physical mechanisms controlling this coupling remain to elucidate.
 
We considered the hypothesis that the shape of the invagination is driven by the flux of protons diffusing on the membrane. Indeed, the absence of cardiolipins in crista membrane induces anomalous crista shapes, and local pH heterogenities induce cristae-like deformations of GUVs comprising cardiolipins \cite{khalifat2008membrane}. 
We described the mechanical properties of mitochondrial crista membrane using a pH-dependent Helfrich model coupled to a diffusive proton concentration field on the surface. We first derived the stationary Green function of this system and showed that it can qualitatively reproduce the bumpy shapes of the mitochondrial invaginations observed experimentally. We then studied the dynamical shape change of the invagination for a membrane tubule subjected to an oscillating proton concentration field, modeling the oscillations between state IV (no ATP synthesis) and state III (high rate of ATP synthesis). We showed that, for appropriate values of the parameters, it can reproduce the salient features of the experimentally-observed shapes of the mitochondrial cristae, oscillating between regular tubes in state IV and bumpy tubular protrusions in state III.

As a next step, for more realism, less symmetric morphologies closer to experimental observations of crista shapes, such as flat balloons with proton sinks on the rim and proton sources on flat zones, could be considered \cite{davies2011}. 

The phenomelogical model introduced here is a coarse-grained description of a membrane containing different lipids, proteins, etc. It will be interesting to consider a more microscopic model of the membrane to gain insight on the values of the parameters $\delta \kappa$, $\delta C_0$, $\delta \sigma$ of this model, and test whether realistic values yield membrane shapes resembling experimental observations.

Finally, a coupling between the shape and the concentration field could be introduced by considering the advection of the protons on the deformed membrane.

In vivo, the model could be tested with super-resolution imaging techniques by changing the energy demand and by monitoring crista morphology variation within mitochondria. Alternatively, experiments on biomimetic membranes can be performed where a local proton flux is introduced, and membrane fluctuations are followed. ATP synthase and complexes of the respiratory chain can be co-reconstituted in liposomes to closely mimic the functioning of the respiratory chain \cite{sjoholm17}. Finally, proton diffusion along membranes mimicking the lipid composition of inner mitochondrial membranes can be estimated using lipid-anchored pH sensor fluorescent dye. Such experiments should bring measurements allowing to test the ability of our model to predict crista morphology.

%\section*{Acknowledgments}
%We thank F. Joubert and S. Bonneau for fruitful discussions. N. P. and H. B. thank the CNRS for the grant XXX.

\appendix
\section*{Appendix}
\subsection{Covariant stress tensor for a membrane}

In this subsection we derive the stress tensor associated with the Hamiltonian 
\begin{align}
H=\int_{\Omega}[\frac{1}{2}\kappa(s)(C-C_0(s))^2+\sigma(s)]dA\label{eq:hamiltonianhelfrich},
\end{align}
 
% for a generic Hamiltonian associated with energy penalties for bending of an arbitrary inhomogeneous surface, 
%\begin{equation}
%H_{\Omega}[g_{ab},K_{ab}]= \int_{\Omega}\mathcal{H} [g_{ab},K_{ab}]dA\label{eq:Hamiltonian_genericaf}.
%\end{equation}
%where $g_{ab}$ and $K_{ab}$ are the metric and the curvature tensors and $dA$ is the area of a surface element. The local force density of a membrane in an equilibrium shape vanishes \cite{AF}, and this force density is the covariant divergence of the local surface stress. 

We consider an infinitesimal deformation of the surface $\mathbf{X}\rightarrow\mathbf{X}+\delta\mathbf{X}$. Instead of tracking the variation of the intrinsic basis and the fundamental forms of the surface induce by this deformation, we enforce these geometrical constraints by introducing Lagrange multipliers and we work with a generalised Hamiltonian $H_c$, 
\begin{widetext}
\begin{eqnarray}
H_c&=&H+\int_{\Omega}\lambda^{ab}(g_{ab}-\mathbf{e}_a\cdot\mathbf{e}_b)dA+\int_{\Omega}\Lambda^{ab}(K_{ab}-\mathbf{e}_a\cdot\nabla_b\mathbf{n})dA+\int_{\Omega}\mathbf{f}^a\cdot(\mathbf{e}_a-\nabla_a\mathbf{X})dA\\&+&\int_{\Omega}\lambda^a_{\bot}(\mathbf{e}_a\cdot\mathbf{n})dA+\int_{\Omega}\lambda_n(\mathbf{n}^2-1)dA.
\label{eq:fullhamiltonianwithlagrangemultipliers}
\end{eqnarray}
\end{widetext}
in which the matrices $\lambda^{ab}$ and $\Lambda^{ab}$ enforce the definition of metric and the curvature, $\mathbf{f}^a$ pins the basis vector to the tangent of the surface, $\lambda^a_{\bot}$ enforces the normal vector to be perpendicular and $\lambda_n$ its normalisation.
The minimisation of the Hamiltonian Eq. (\ref{eq:fullhamiltonianwithlagrangemultipliers}) with respect to the, now, 12 independent functions give the following equations,
\begin{eqnarray}
\label{eqLM1}
 \frac{\delta H_c}{\delta {\bf X}} &=& \nabla_a\mathbf{f}^a  \\
 \label{eqLM2}
 \frac{\delta H_c}{\delta \mathbf{e}_a} &=&- \mathbf{f}^a   +   ( \Lambda^{ac}K^b_c+2\lambda^{ab})\mathbf{e}_b    -\lambda^a_{\bot}\mathbf{n}\\
   \frac{\delta H_c}{\delta \mathbf{n}}&=&( \nabla_b\Lambda^{ab}+\lambda^a_{\bot})\mathbf{e}_a          + (2\lambda_n-\Lambda^{ab}K_{ab})\mathbf{n}\\
\frac{\delta H_c}{\delta K_{ab}}&=& \Lambda_{ab}  + \mathcal{H}^{ab}\\
\label{eqLM5}
 \frac{\delta H_c}{\delta g_{ab}}&=& \lambda^{ab} + \lambda_{\rho}\rho\sqrt{g}\frac{g^{ab}}{2}   + \lambda_{\phi} \rho\phi\sqrt{g}\frac{g^{ab}}{2} -\frac{1}{2}T^{ab}
\label{eq:EL_equations}
\end{eqnarray}
where $\mathcal{H}^{ab}$ and $T^{ab}$ are defined as the functional derivatives of the Hamiltonian density, ${\mathcal H}=\frac{1}{2}\kappa(s)(C-C_0(s))^2+\sigma(s)$ with respect to the two fundamental forms
\begin{eqnarray}
\mathcal{H}^{ab}&=&\frac{\delta\mathcal{H}}{\delta K_{ab}}\\
T^{ab}&=&-\frac{2}{\sqrt{g}}\frac{\delta \sqrt{g}\mathcal{H}}{\delta g_{ab}}.
\end{eqnarray}

 We notice that the Lagrange multiplier $\mathbf{f}^a$  obeys a conservation law $(\nabla_a\mathbf{f}^a=0)$. Using that Eqs. (\ref{eqLM1}-\ref{eqLM5}) vanish, we eliminate the Lagrange multipliers present in Eq. (\ref{eqLM2}) and find 
\begin{equation}
\mathbf{f}^a=(T^{ab}-\mathcal{H}^{ac}K^b_c)\mathbf{e}_b-(\nabla_b\mathcal{H}^{ab})\mathbf{n}.\label{eq:stresstensor_result}
\end{equation}
The quantity $\mathbf{f}^a$  can be identified as the surface stress tensor\cite{Guv04a,Deserno_curvatureconvention}. The surface stress tensor is a 2x3 matrix and its coordinates in the intrinsic basis of the surface can be written as
\begin{eqnarray}
\label{fabHapp}
\mathbf{f}_H^{ab}&=T^{ab}-\mathcal{H}^{ac}K^b_c \\ \label{fanH}
f_H^{an}&=-(\nabla_b\mathcal{H}^{ab}).
\end{eqnarray}
 
%Results from \cite{Muller_phd} for simplified expressions of $T^{ab}$ and $\mathcal{H}^{ab}$ yield

%\begin{eqnarray}
%T^{ab}&=-\mathcal{H}g^{ab}+2\frac{\partial \mathcal{H}}{\partial C}K^{ab}+2\frac{\partial\mathcal{H}}{\partial\mathcal{R}}\mathcal{R}g^{ab}\\
%\mathcal{H}^{ab}&=\frac{\partial\mathcal{H}}{\partial C}g^{ab}+2\frac{\partial\mathcal{H}}{\partial\mathcal{R}}\cdot(Cg^{ab}-K^{ab})\label{eq:expressionsforTabHab}
%\end{eqnarray}
%where we recall that $C$ and $\mathcal{R}$ are the invariant quantities composed from the principal curvatures and are defined in equations (\ref{eq:defn_C}) and (\ref{eq:defn_R}).
%Inserting these expressions in  Eq.  \ref{eq:stresstensor_result} we have
%\begin{widetext}
% \begin{eqnarray}
% \mathbf{f}^{a}= \left(-\mathcal{H}+\rho\frac{\delta\mathcal{H}}{\delta\rho}+2\frac{\partial\mathcal{H}}{\partial\mathcal{R}}\mathcal{R} \right) g^{ab}\mathbf{e}_b+\left(\frac{\partial\mathcal{H}}{\partial C}-2\frac{\partial\mathcal{H}}{\partial\mathcal{R}}C\right)K^{ab}\mathbf{e}_b+2\frac{\partial\mathcal{H}}{\partial\mathcal{R}}K^{ac}K^b_c\mathbf{e}_b-\nabla_b\mathcal{H}^{ab}\mathbf{n}\label{eq:stresstensorfullexpression}
% \end{eqnarray}
%\end{widetext}
%Equation (\ref{eq:stresstensorfullexpression}) is our most important formal result where we have incorporated inhomogeneous fields with conservation laws into the expression for what we will soon see is the stress tensor in the most generic case imaginable.

\subsection{Coefficients for dynamical force balance equations\label{app:coeffsdyn}}

The full expression for the dynamical force balance equations Eqs. (\ref{forcetangentcoef}),(\ref{forcenormcoeff}) can be obtained by inserting the coefficients which are given as follows:

\begin{subequations}
\begin{eqnarray}
       a_1&=&\frac{\eta_b-\eta_s}{R}\\
    a_2&=&\eta_b+\eta_s\\
    a_3&=&-\frac{(1-X)^2}{2R^2}\delta\kappa-\delta\sigma+\frac{2R\sigma_0}{1+X}\delta C_0\\
 b_1&=&\frac{\eta_b+\eta_s}{R^2}\\
 b_2&=&\frac{\eta_b-\eta_s}{R}\\
 b_3&=&\frac{2\sigma_0}{R^2(1-X^2)}\\
 b_4&=&\frac{4X\sigma_0}{1-X^2}\\
    b_5&=&\frac{2R^2\sigma_0}{1-X^2}\\
    b_6&=&-\frac{(1-X^2)}{2R^3}\delta\kappa+\frac{1}{R}\delta\sigma+\frac{2X\sigma_0}{1-X^2}\delta C_0\\
    b_7&=&\frac{1}{R(1-X^2)}\Big(2R^3\delta C_0\sigma_0\nonumber\\ &-&(1-X)^2(1+X)\delta\kappa\Big)
\end{eqnarray}
\end{subequations}
Here $\kappa_0$ has already been substituted in terms of $\sigma_0,R$ and $X$ using the relation between the Helfrich parameters (\ref{eq:relationbetweenparameters}). If we set $a_1,a_2,b_1$, and $b_2$ to zero (i.e. no viscosities), we obtain the steady state force balance equations for the Helfrich cylinder model.

\subsection{Details of the numerical resolution of hydrodynamics equations}
The system of equations Eqs. (\ref{eq:forcebalancesnumerics1},\ref{eq:forcebalancesnumerics2}) is solved numerically at time $t$ using the command NDSolve of the Mathematica software. The solutions obtained for $v_s^t$ and $v_n^t$ are discretized in space on a regular grid with a step equal to 10$^{-2}$ for a cylinder with radius $R=1$. The discrete spatial derivatives of the velocities $(v_n^{(i),vect}, (i=1,..4), v_s^{(j),vect}, (j=1,2)$  are derived on this grid.
 
The vectors are interpolated using polynomials of degree 18 to obtain analytical functions $(\partial^i_s v_n^{pol}, \partial^j_s v_s^{pol}, i=1, ...,4, j=1, ...,2)$ that are used to derive the deformation fields and their derivatives at the time step $t$,
\begin{eqnarray}
\partial_s^i u_s^{t}(s)=\partial_s^i u_s^{t-dt}+dt\times \partial_s^i v_s^{pol}, \quad i=1, ...,2\\
\partial_s^j u_n^{t}(s)=\partial_s^j u_n^{t-dt}+dt\times \partial_s^j v_n^{pol}, \quad j=1, ...,4.
\end{eqnarray}

\bibliographystyle{unsrt}
\bibliography{nirbhaybib}
   
\end{document}